# Engineering and performance of a high-resolution and long stroke linear slider for cryogenic environments based on superconducting magnetic levitation

Ignacio Valiente-Blanco*, Efren Diez-Jimenez and Jose-Luis Perez-Diaz

*Dpto. de Ingeniería Mecánica, Universidad Carlos III de Madrid, Butarque, 15, E-28911 Leganés, Spain*

## Abstract

In this paper, a contactless linear slider for precision positioning able to operate in cryogenic environments is presented. The device, based on superconducting magnetic levitation, does not present contact between the slider (composed of a permanent magnet) and the guideline (made of high-temperature superconducting disks) of the mechanism, thereby avoiding any tribological problems. Moreover, the slider is self-stable and the superconductors provide inherent guidance to the permanent magnet in the sliding DoF due to the high translational symmetry of the magnetic field that leads to low power consumption. A sub-micrometre resolution and a symmetric stroke over ±9 mm have been demonstrated at cryogenic temperatures. In addition, a set of design rules for this kind of mechanism has been proposed and experimentally validated. These rules demonstrate that the performance of the device can be tuned just by modifying some geometrical parameters of the mechanism. In this way, the sensitivity and stiffness, resolution, angular run outs and power consumption can be adjusted for different applications and requirements.

* Corresponding author: Ignacio Valiente Blanco  Address: Avenida de la Universidad 30, Universidad Carlos III de Madrid, Spain
Tel.: +34 91624998380; fax: +34 916249912.
**E-mail address: ivalient@ing.uc3m.es**

# 1. Introduction

Nanotechnology and nanoscience have experimented an great growth in the last decades [1,2]The motivation for this has been the increasing demands of microelectronics, optics, biomedical and precision engineering [3]. Ultra precision manufacturing and inspection systems in micro-automating semiconductor fabrication, copying machines, stepper stages for photolithography, small-scale measuring machines (CMMs) for large area scanning, surface imaging in scanning probe microscopy (SPM), nanopositioning and nanomeasuring machines (NPM-Machine), development of micro-assemblies and micro and nanoelectromechanical systems (MEMS and NEMS) and servo systems of hard-disk drives (HDD), are just a few examples of the wide range of applications where micrometer/nanometer positioning within a long range is required [4–6].

Additionally, in some applications, cryogenic environments (temperatures below 120 K) are a desirable or mandatory condition. The sensitivity of a large number of sensors is greatly increased when they are at cryogenics temperatures, like for example, those required for far infrared interferometer spectroscopy [7,8]. The operating conditions in these environments include very low temperatures but also high vacuum. In this context, it is challenging for these mechanisms to overcome all the tribological problems associated with these conditions [9,10]. In addition very low energy consumption is also desirable in cryogenic environments.

Different technologies have been applied to precise positioning. Usually, piezoelectric (PZT) actuators are chosen to design precise positioning systems in order to achieve the submicrometric or nanometric positioning. Despite that the main limitation of PZT actuators is their very limited motion range, usually not longer than a few hundred micrometers [11,12], hybrid or dual-stage positioning systems based on piezoelectric actuators and friction type piezoelectric actuators (based on the slip-stick phenomenon [13]) can sometimes reach long strokes with impressive accuracy [14]. Nevertheless piezoelectric actuators have several limitations, for example, they are sensitive to environmental changes such as temperature [15]. Nonlinearities such as hysteresis and creep are also present in piezoelectric materials [16,17]. Furthermore, the voltage required to operate piezoelectric actuators can be as high as several-hundred volts, which can be a problem in situations where it is necessary to have low levels of electrical input.

Other researchers have focused their efforts on magnetic levitation (maglev) positioners. These devices eliminate the strain, backlash, and hysteresis that limited the precision of position control and by being contactless devices, long lifetime and fatigue problems are minimized. Several prototypes have been proposed with strokes of tens of mm and resolution in the nanometer scale in one or more DoF [18–23]. In comparison with the piezoelectric positioners, the active magnetic levitation systems can offer similar accuracies for longer motion ranges. Nevertheless, the main disadvantage of these active magnetic levitation devices is that a great effort is required to control them as they are naturally unstable [24]. This requires a complex control method and consequently, complex electronics. In addition, the active control generates extra heat and requires high currents in the electromagnets [25]. Both, the need of high currents and generation of heat, are a main issue in cryogenic environments.

Other devices based on giant magnetostrictive materials has been proposed [26–28]. The current demanded in these devices is high (sometimes over 10 A), and strokes and resolution are very limited and very temperature dependent.

Devices based on superconducting magnetic levitation (SML) seem to be a very suitable option for actuators and positioners in cryogenic environments. SML provides self-stable levitation of a permanent magnet (PM) over a high temperature superconductor (HTS) [29,30]. At cryogenic temperatures HTS are naturally in the superconducting state and no cooling power is required. However, only little attention has been paid in positioners based on this technology. Some kind of conveyors has shown reasonable results with strokes of not more than a few mm and resolutions in the µm range [31,32].

The invention presented in this paper [33] takes advantage of superconducting magnetic levitation technology to obtain long stroke positioning with high precision, minimized run-outs, at cryogenic temperature operation and low power consumption.

In addition, the lack of contact between the moving parts is very suitable for operation in clean-room applications too, such as in the semiconductor manufacturing industry [34]. The device is based on the fact that a contactless sliding kinematic pair is established between a long permanent magnet levitating over a HTS at the mixed state. The sliding DoF (X direction) is generated thanks to a high translational symmetry of the applied magnetic field on the superconductor within the stroke of the mechanism. However, the choice of shape, configuration and size of the superconductors and magnets to determine the mechanical behavior is not a small issue and require analytical and experimental studies [35–40]. A initial mechanical characterization of the linear slider has been presented in [35].

In this paper, the working principles of the linear slider are presented, theoretically analyzed and experimentally validated. A sub-micrometer resolution and a symmetric stroke over ± 9 mm have been demonstrated at cryogenic temperatures. Furthermore, the theoretical foundations provide several design rules that can help to adapt the device for different sought requirements of stiffness, position resolution, run outs and power consumption. The mentioned design rules have been experimentally validated cooling down the superconductors with $LN_2$ at 77 K.

## 2. Description of the device

The linear slider can be divided in two sub-systems: the guidance system and the actuating system. The guidance system is composed of a static guideline, made of two 45 mm diameter superconducting polycrystalline $YBa_2Cu_3O_{7-x}$ disks (HTS) (1), and a slider composed of a long $Nd_2Fe_{14}B$ permanent magnet (PM) (2). The PM is a bar 160 mm in length and a square section of 10x10 $mm^2$ with a remanence $Br = 1.3T$ and a coercivity $Hc = 900$ kA/m. Notice that its magnetization direction is parallel to the Z axis in Fig 1. The rest of the mechanical fixtures are made of aluminum 6061 to avoid magnetic effects. The actuating system is composed of the same PM and a pair of coils (3) placed at both ends of the stroke of the mechanism. The axis of the coils is parallel to the Z axis and contained in the XZ plane in Fig 1. These coils are specially designed to provide improved stability to the PM and only exert magnetic forces in the sliding direction (X direction). They are manufactured with an inner aluminum core of $80 \times 30$ $mm^2$ and 10 mm height surrounded by 150 turns of wire.

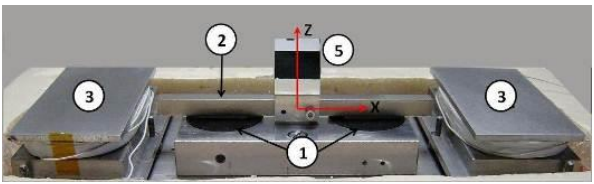

**Fig. 1**. Figure of the linear slider: 1 YBaCuO superconductor disks; 2 Slider permanent magnet; 3 Coils, 5 polished aluminum mirror cube.

## 3. Theoretical foundations

### 3.1 Working principle

Once the superconductors are at a temperature below their critical temperature (~90 K), the slider levitates stably and without contact over the superconductors in a centered equilibrium position. The superconductors not only provide the lift required but also the guidance of the permanent magnet in its path along the sliding direction X.

When the magnet is levitating it can be moved just by circulating a current in the actuating coils. These two coils can exert an attraction–repulsion (pulling-pushing force) on the PM, depending on the direction of the circulating currents. With these coils, the equilibrium point of the slider can be modified. Due to the naturally stable levitation and guidance provided by the HTS, only the control of the position in the sliding direction is required. This paper is mainly focused on the mechanical behaviour of the slider, hence a simple open-loop control strategy is used.

In order to obtain guidance of the slider, the magnetization direction of the PM and the coils must be carefully designed. Due to the translational symmetry of the magnetic field generated by the PM, a sliding kinematic pair is established between the PM and the superconductors in the sliding direction. Thus the slider can be moved in the sliding DoF with very low resistance. On the contrary, greater restoring forces appear if the PM is moved in any other direction.

When the PM is moved away from the initial equilibrium position in the sliding direction, the small restoring force in the sliding direction exerted on the superconductor can be calculated using the following equation:

$$F_{SC_X} = \int_{SC} M_x \cdot \left(\frac{\partial B_x}{\partial X} + \frac{\partial B_x}{\partial Z}\right) + B_x \cdot \left(\frac{\partial M_x}{\partial X} + \frac{\partial M_x}{\partial Z}\right) dV$$

**Eq. 1.**

where Mx is the magnetization of each differential element of the superconductor in the X direction, Bx, is the magnetic flux density on each differential element of the superconductor in the X direction, and V stands for volume of the superconductor.

If the PM is moved away the initial cooling position, the symmetry of the magnetic field applied is broken (due to boundary effect of the flux lines in the ends of the PM) and forces appear trying to restore the initial equilibrium position. However, due to the fact that the magnetic field in the X direction applied on the superconductors is small (various orders of magnitude lower than the remanence of the PM), these restoring forces are smaller than those that appear if the PM is moved in any other direction. The magnetic flux density on the superconductors at the normal state and the magnetic flux in the sliding direction for three positions of the stroke of the mechanism are shown in Fig 2.

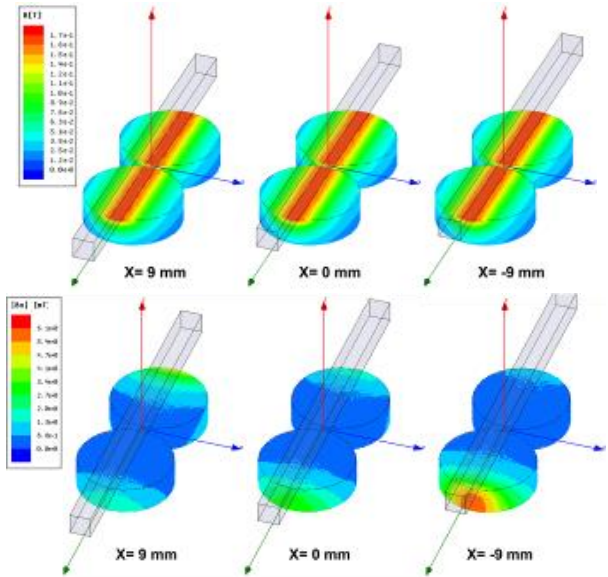

**Fig 2.** Total Magnetic flux (up) and magnetic flux in the sliding direction (down) for three positions of the stroke of the mechanism: equilibrium position and end-stroke positions.

## 3.2 Design rules

The previous magnetic analysis depends on the geometry of each component. The total stroke of the mechanism is influenced by the size of the PM and the HTS, their relative position and the separation between the coils that establish the limit of the displacement. Then, the stroke of the mechanism can be modified by modifying the dimensions of the prototype. In addition, some other parameters of the performance can be tuned too.

### 3.2.1 Sensitivity and stiffness in the sliding DoF

The magnetic field gradient and the magnitude of the magnetic field affect the stiffness (restoring forces divided by X displacement) and sensitivity (X displacement divided by current in the coils) of the slider. It has been reported that the magnetic stiffness between a PM and a HTS at the mixed state is increased for lower values of the *HFC* [42–44]. The higher the magnetic gradient and the higher the magnetic field strength, the higher the required acting force and the stiffness in the sliding direction. Hence, run outs can be reduced if an enhanced magnetic stiffness is achieved. Therefore, the stiffness, sensitivity and run outs of the slider can be tuned by modifying two parameters: the distance between the superconductors *d* and the Height of Field Cooling (*HFC*), see Fig 3.

The sensitivity and stiffness are also related to the position resolution of the mechanism in the sense the higher the stiffness, the lower the sensitivity and therefore, a lower current resolution is required to achieve the same positioning resolution.

### 3.2.2 Current requirement and power consumption and power consumption

It has to be taken into account that the foregoing modifications will also affect the current requirement in the coils and the power consumption of the mechanism. Then, a higher power consumption is expected if *HFC* is reduced and if the distance between the superconductors, *d*, is increased.

### 3.2.3 Position resolution

Position resolution can be improved if the sensitivity of the device is reduced. However, this will bear a rise in the power consumption. A compromise between sensitivity and maximum current shall be achieved. Separately, in order to improve the position resolution an improvement in the current resolution in the coils is always desirable.

### 3.2.4 Run outs

In a previous work, it has been established that angular and lateral run outs (except from the pitch) are mainly due to misalignments between the PM magnetization axis and the magnetic axis in the coils [45].

In this previous work, the greater run out is the pitch (rotation around Y direction), which is also critical for the straightness of the motion. Such a great pitch is mainly caused by the loss of lift and unbalance of the PM when it is moved away the initial equilibrium position. In the present work the pitch is deeply analyzed and corrected. A simplified mechanical model for the pitch of the slider is proposed in Fig 3.

In this model the following simplifications have been considered: the PM is considered as a rigid body, self-alignment of the PM with the magnetic field generated by the coils is neglected due to small pitch angles involved, and it is also neglected the interaction between the coils and the superconductors and displacements of the center of masses of the PM in Z axis (see Fig. 3) are neglected. With the foregoing hypothesis in mind, forces between the PM and two superconductors at the mixed state located at distance *d* between them are approximated to the equivalent forces exerted by two springs of constant stiffness $K_z$, and initial length of the spring $Z_0 = HFC$ as show in Fig 3.

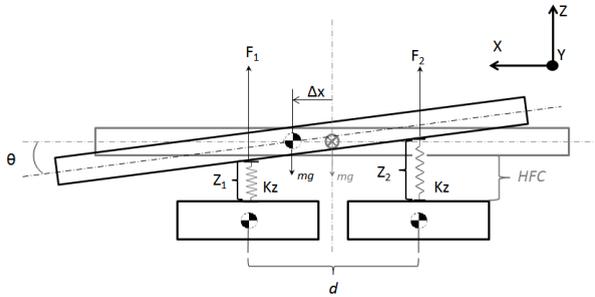

**Fig 3.** Mechanical simplified model scheme.

It can be proved that, pitch of the slider in the foregoing model is described by the following equation:

$$\theta = \frac{2mg}{K_z d^2} X$$

Eq. 2.

where
m is the mass of the slider (170 ±5 g),
g is the gravity acceleration,
Kz is the stiffness of the equivalent springs in Fig. 3.
*d* is the distance between the superconductors
and X is the position of the slider.

According to the proposed equation, a few additional rules can be derived. Pitch is linearly dependent of the displacement of the slider, pitch can be quadratically reduced if the distance between the superconductors is increased and pitch is linearly dependent on the mass of the slider.

In addition, the pitch of the slider can be modified by magnetic self-alignment of the PM and the magnetic field generated by the coils. The alignment torque of the PM immersed in the magnetic field generated by the coils can be calculated as:

$$T = \int_V [r \times (M \cdot \nabla)B + M \times B] \, dV$$

Eq. 3

where,
r is the position vector of each dipole in the magnet,
M is the magnetization of the PM (considered constant),
B is the magnetic flux density
and V is the volume of the PM.

Therefore, if the relative position between the coils and the PM is modified, this alignment torque can be used to our benefit. Specifically, a reduction of the pitch can be achieved by elevating the coils with respect to the axis of the PM (parameter *Hc*).

### 3.4 Decoupled model: FEM calculations

Calculation of forces is not an small issue, and FEM calculations usually require to develop complex models with limited application and accuracy [46] that demand high computational and time efforts. In order to obtain the force exerted on the PM, a decoupled model is proposed. In this model, the presence of the superconductors is neglected when calculating the actuating force exerted by the coils on the PM. Under this hypothesis, the force that the superconductors exert on the PM is equivalent but opposite directed to the force exerted by the coils on the PM for each equilibrium point of the slider in its path. A diagram of the decoupled model implemented in a FEM software is shown in Fig. 4. The proposed FEM model has been used to estimate the stiffness of the device in the sliding DoF.

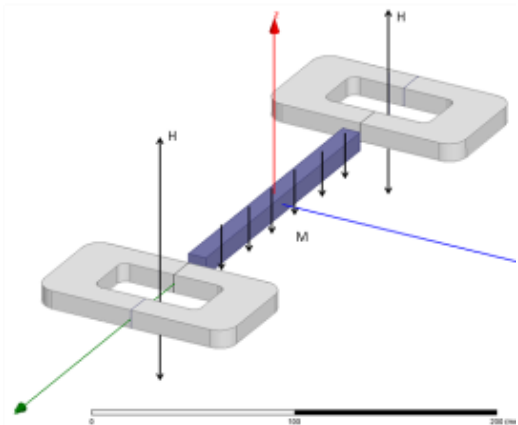

**Fig 4.** Decoupled magnetomechanical FEM model of the actuating system.

## 4. Experimental set-up and results

A campaign of experiments has been carried out in order to validate the design rules and prove the position resolution. The details of these experiments are reported in this section.

### 4.1 Experimental set up and procedure

The experiments have been developed using the experimental set-up presented in Fig 5. The prototype has been contained in a thermal isolating vessel (8) and immersed in a liquid nitrogen bath at ambient pressure (~77 K). The height of the prototype with respect to the beam from the interferometer/autocollimator can be regulated thanks to a lab-jack stand (6). Everything has been assembled onto an optic table (7).

Position resolution and accuracy of the slider have been precisely measured using an Agilent 10706B High Stability Plane Mirror Interferometer with 0.62 nm accuracy. However, for some measurements in where a good precision in the position is not needed a common scale with an accuracy of 250 µm have been used. The slider angular run outs have been measured using an Newport LDS-Vector auto-collimator with an accuracy of 3% of the measurement.

From the design rules, there are 3 main geometric parameters than affect to the performance of the slider:

- The height of levitation or height of field cooling (*HFC*), which is the distance from the bottom surface of the PM and the top surface of the HTS, as shown in Fig 5.
- The distance between the superconductors (*d*), understood as the distance between the axis of the HTS in the sliding direction of the mechanism.
- The height of the coils (*Hc*), which is the vertical distance (Z axis) between the imaginary lines that passes through both the centers of the coils and the axis of the PM.

This geometric parameter has been set up with an accuracy of 0.1 mm.

The stroke is divided in two ranges from X=[-9,0] mm and from X=[0,+9] mm. The slider is moved along each range just by one coil while the other is switched off. The current in the actuating coil has a circulating direction that repels the PM (pushing force).

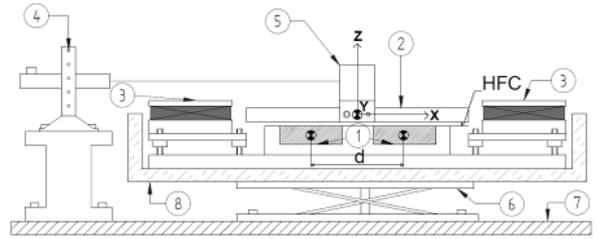

**Fig. 5**. Sketch of the experimental set-up: 1 YBaCuO superconductor disks; 2 Slider permanent magnet; 3 Coils, 4 Interferometer/Auto collimator; 5 polished aluminum mirror cube; 6 lab-jack stand; 7 optic table and 8 liquid nitrogen vessel. *d*: distance between the superconducting disks and *HFC*= height of field cooling.

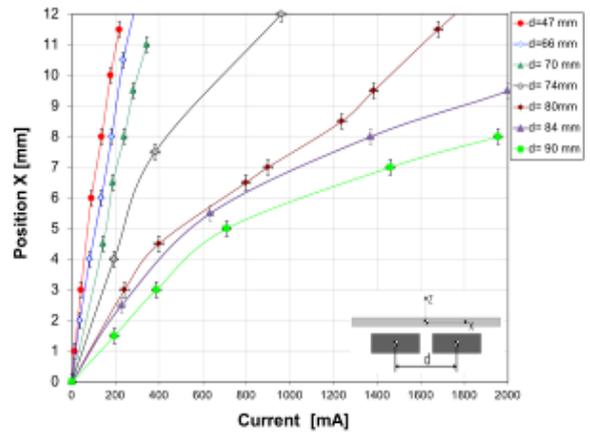

**Fig 6.** Measured X position vs. current in the coil for different distance between the superconducting disks (*d*). *HFC*= 3mm

The current in the coils has been supplied from a power source with two independent channels and measured with two 10 µA resolution multimeters. Once the levitation is established, the current in the corresponding coil is modified and the position, the run outs and the current in the coils measured for different currents. Finally, an optic mirror cube (5) made of polished aluminum and one cubic inch in size was fixed to the slider (2) to reflect the laser beam from the interferometer/auto-collimator.

### 4.2 Experimental results and discussion

#### 4.2.1 Sensitivity and stiffness of the slider

In Fig 6, X position of the slider vs. current in the coil is plotted for different values of *d*. From this figure it is clear that the highest the value of *d,* the lower the sensitivity.

The acting force on the PM has been calculated using the decoupled model introduced in section 3. The results of FEM calculations of forces in the sliding direction are plotted in Fig. 7 for different values of the distance between the superconductors.

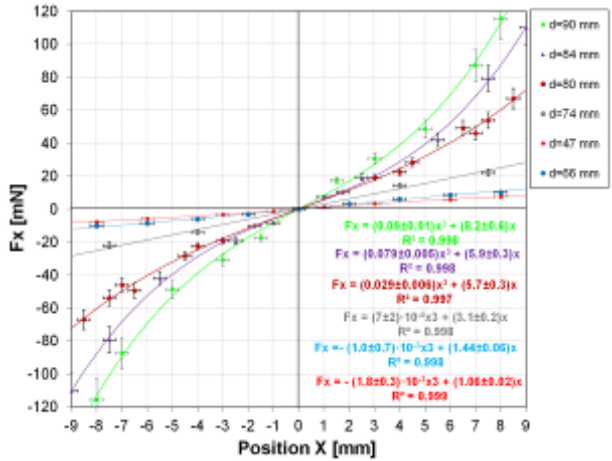

**Fig 7.** Calculated forces vs. X position for different distance between the superconductor disks (*d*). *HFC*= 3mm.

The sign of the forces and displacements in Fig. 7 shows that the origin of coordinates is a stable equilibrium point in the stroke of the slider. It is also relevant that the function Fx vs. X position of the slider fits a cubic polynomial with an $R^2 \geq 0.997$ in all cases. As it can be seen, the required actuating force changes considerably with respect the value of *d*. Therefore, stiffness and sensitivity are very dependent on the value of *d*. Deriving the fitting curves the stiffness in the sliding DoF can be calculated. In Fig.8, stiffness vs. X position of the slider is plotted for different values of *d*.

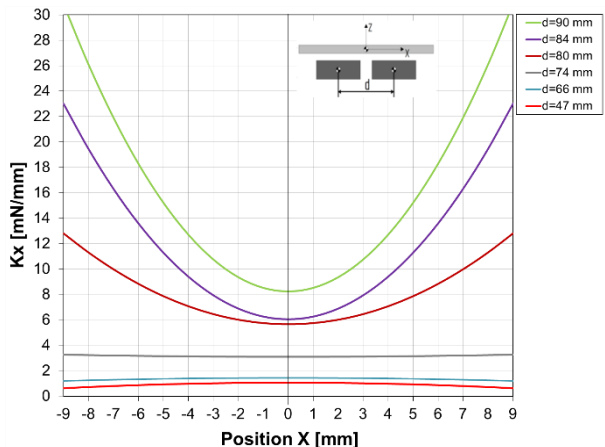

**Fig 8.** Calculated stiffness vs. X position of the slider for different values of *d*. *HFC*= 3 mm and Hc= 0 mm in all cases.

A similar behaviour can be expected when modifying the *HFC*. X position of the slider vs. current in the coils is plotted in Fig. 9 for different values of *HFC*. It can be observed that when the *HFC* is reduced the current increases. This can be explained by the increase in the magnetic field strength and gradient on the superconductor as a consequence of reducing the distance between the PM and the HTS.

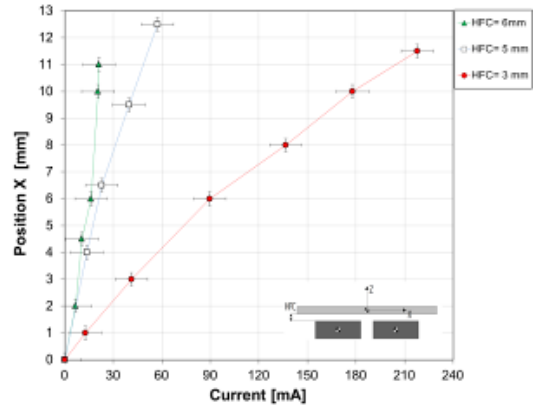

**Fig 9.** Measured position X vs. current in the coil for different values of *HFC*. *d*=47 mm..

### 4.2.2 Current requirement and power consumption

If the current required to achieve a certain stroke is increased, the current requirement in the coils will increase. The magnitude of the current in the coils vs. d to achieve a stroke of±9 mm is plotted in Fig. 10. For values of *d* higher than 70 mm, For values of d higher than 70 mm, there is a large increase in the current needed in the coils.

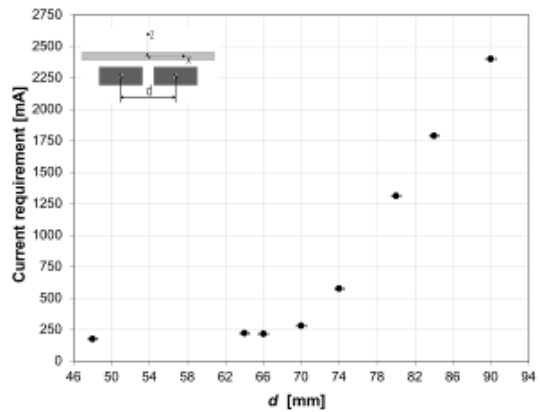

**Fig 10.** Current requirement vs. distance between superconductors. *HFC*= 3 mm and Hc= 0 mm in all cases.

The demanded current in the coils is also increased when *HFC* is reduced, Fig 11.

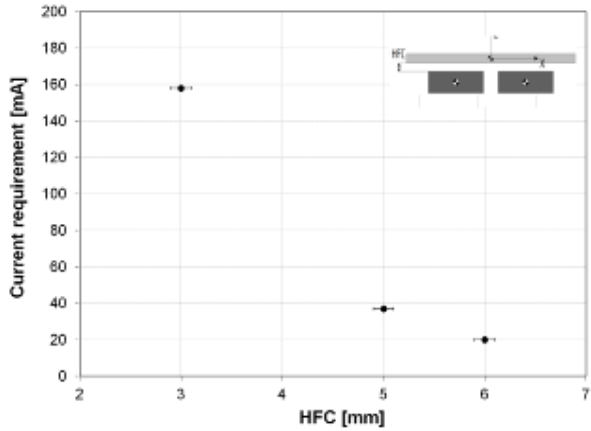

**Fig 11.** Current requirement vs. *HFC*. *d*= 47 mm and Hc= 0 mm in all cases.

The power consumption of the device is difficult to define insofar as it is very much influenced by the operation conditions and the environment. For example, the power consumption due to the resistance of the wires in the coils is very dependent on the working temperature and on the residual resistance ratio (RRR) of the copper in the wires [47]. Then, a working temperature must be defined. However, it is clear that the higher the HFC and the lower the value of d, the lower the power consumption, regardless of the working temperature.

When one pushing coil is actuating, the maximum power dissipated in the coils is about 16mW for d = 47 mm and HFC = 3 mm and around 2W if d = 90 mm. HFC = 3 mm and RRR = 300 have been considered for previous calculations. However, this power consumption can be drastically reduced by improving the RRR of the wires or by changing the sense of the circulating current (pulling force). In Fig. 12, the X position of the slider vs. current in the coils is represented for a pushing and pulling coil configuration. A clear reduction of the current required in the coil can be observed. However, this change implies an increase in the sensitivity.

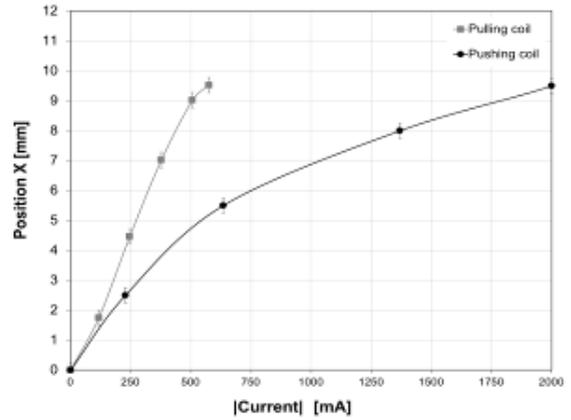

**Fig 12.** X position of the slider vs. magnitude of the current in a pushing and a pulling coil. *HFC*= 3 mm, *d*= 84 mm and Hc= 0 mm in both cases.

### 4.2.3 Position resolution and accuracy

All the positions presented in this section have been measured with an Agilent 10706B High Stability Plane Mirror Interferometer with 0.62 nm accuracy. Position resolution and accuracy are two main issues in high-precision positioners. In Fig. 13, the stability of the slider (open-loop control) is represented vs. time for a prototype with *HFC*= 3 mm, Hc= 5 mm and *d*= 84 mm. The accuracy of the mechanism is of the order of ± 0.3 µm.

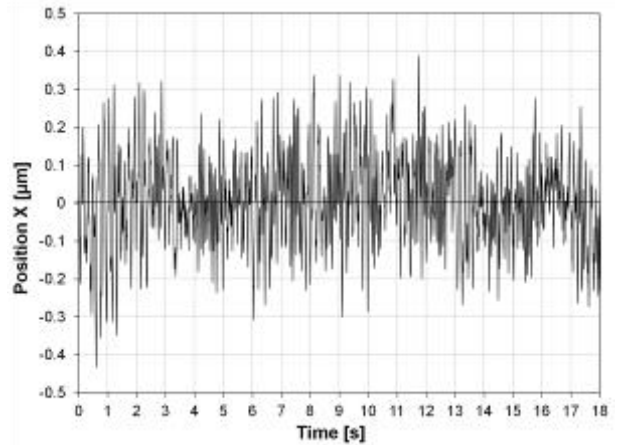

**Fig 13.** Stability of the position along time. *HFC*= 3 mm, *d*= 84 mm and Hc=5 mm in both cases.

A RMS position resolution of 230±30 nm is demonstrated in Fig 14. These measurement have been taken in the surrounding of the initial equilibrium position (where the sensitivity is the maximum) with a current resolution of 15±1 µA . According to results in the previous section, position resolution can be improved if, for example, the *HFC* is reduced, or the value of *d* is increased (provided the same current resolution).

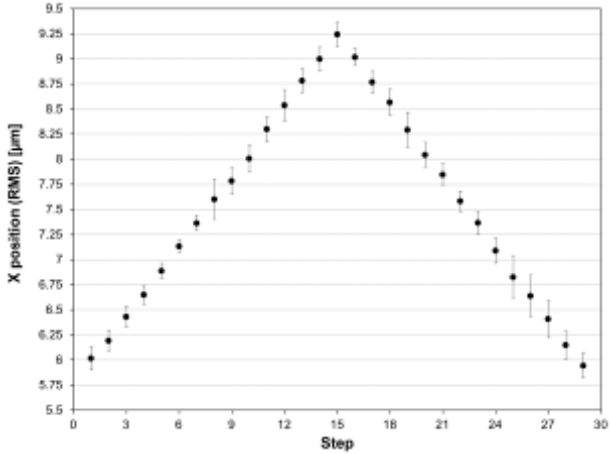

**Fig 14.** X position of the slider vs. current step.

The position resolution could be also improved if the current resolution in the coils is improved. Position resolution vs. current resolution is plotted in Fig.15.

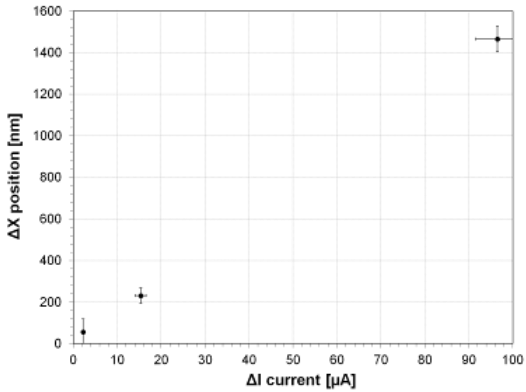

**Fig 15.** Position resolution vs. current resolution.

*4.2.3 Angular run outs*

*4.2.3.1 Pitch*

Pitch of the slider vs. X position for different values of *d* is shown in Fig. 16. A good linearity of the results can be observed in good agreement with eq. 2. Pitch is also reduced when the *HFC* is decreased due to the increased stiffness, Fig 17.

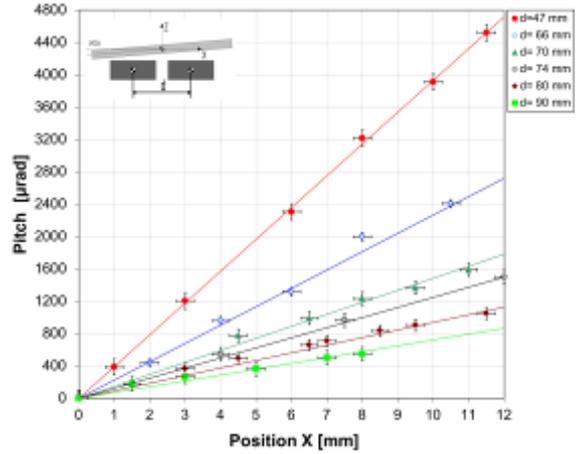

**Fig 16.** Pitch vs. X position of the slider for different values of *d*. *HFC*= 3 mm.

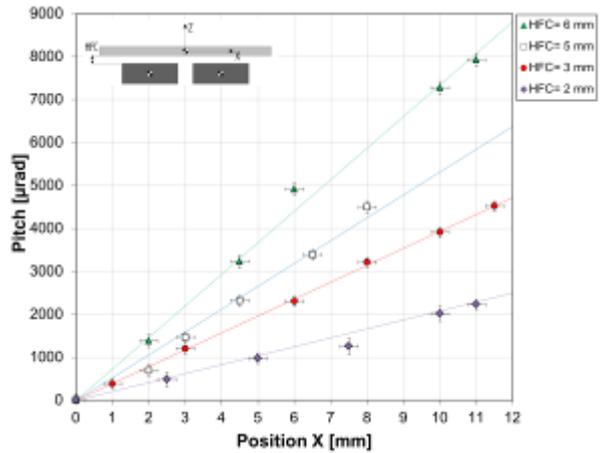

**Fig 17.** Pitch vs. X position of the slider for different values of *HFC*. *d*=47 mm.

The model for the pitch presented in section 3.2.4 has been empirically improved by including a linear dependence of the stiffness $K_z$ with the distance between the superconductors d. The stiffness with respect to *d* is described in eq. 3:

$$K_z(d) = K_{47} + 0.01 \cdot d.$$

where,
$K_{47}$ is the stiffness for *d*=47 mm.

Comparison of the models for the pitch considering this empirically stiffness correction is shown in Fig. 18.

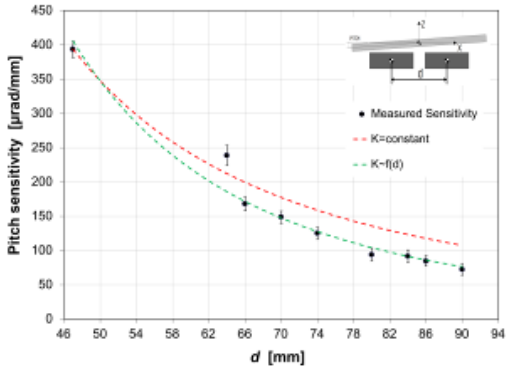

**Fig 18.** Pitch sensitivity vs. distance between the superconductors. Comparison of the models for the pitch..

Moreover, the pitch can be corrected by modifying $Hc$ as introduced in section 3.2.4. Maximum pitch for a 9 mm stroke vs. elevation of the coils $(Hc)$ is plotted in Fig. 19.

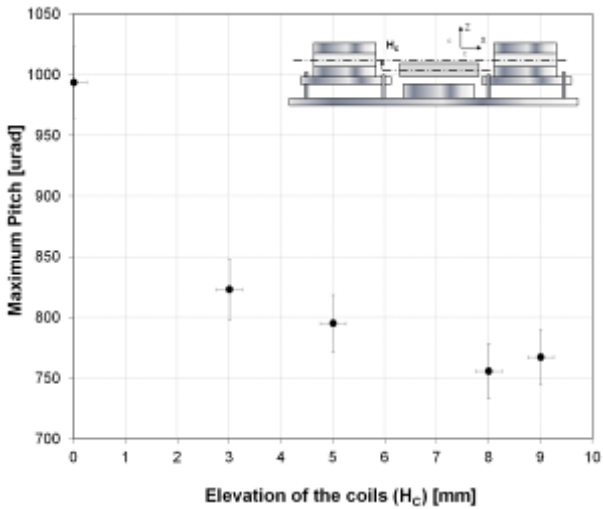

**Fig 19.** Maximum pitch for a 9 mm stroke vs. elevation of the coils (Hc). *HFC*= 3 mm and *d*= 84 mm.

In summary, it can be demonstrated that for values of Hc between 0 and 8 mm the pitch of the slider can be reduced more than a 20%.

*4.2.3.1 Yaw*

All run outs are expected to be reduced if the *HFC* is reduced due to the increased stiffness between the PM and the HTS. Yaw (rotation around Z axis) vs. X position of the slider for different values of *HFC* is plotted in Fig. 20. A reduction about 50 % was observed for a stroke about 11 mm for *d*= 47 mm and *Hc*=0 in all cases.

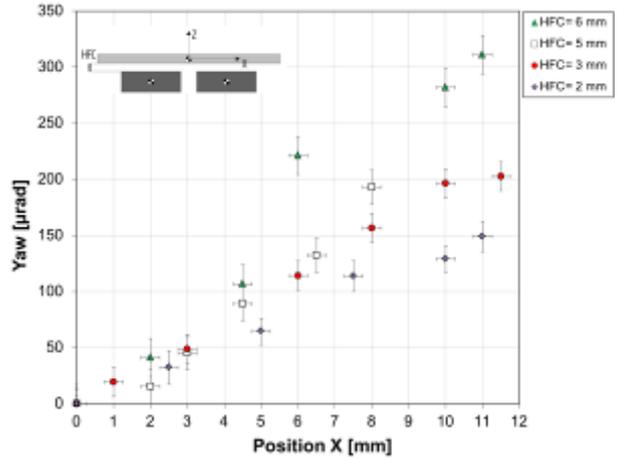

**Fig 20.** Yaw vs. X position of the slider for different values of *HFC*. *d*= 47 mm and *Hc*=0 mm in all cases.

No relationship between *d* and Yaw has been observed in the experiments.

## 4. Conclusions

A high-resolution positioning linear slider with a long stroke has been designed, built and tested for cryogenic applications. The device is mainly composed of two superconducting disks as the stator and a long permanent magnet as the slider. Due to the high translational symmetry of the magnetic field of the superconductors, a contactless sliding kinematic pair is established between them and the permanent magnet. The actuating system is composed of two coils. Then, by regulating the current in the coils, the position of the slider can be modified. A high resolution (230±30 nm) in a long stroke (over ±9 mm) has been demonstrated at cryogenic temperatures (77 K).

Besides, a set of design rules has been proposed and experimentally verified. Following these design rules, several parameters of the performance of the device can be tuned. These rules are summarized below:

- The sensitivity, the stiffness and the resolution can be tuned by modifying the *HFC* or the distance between the superconductors *d*.
- The resolution of the device can be improved linearly improving the current resolution.
- The power consumption can be reduced by more than one order of magnitude with an anti-serial connection of the coils in the actuating system.
- The power consumption can be reduced if higher quality wires are used in the coils (lower RRR).

- Angular run outs can be reduced by more than 200% by reducing the HFC.
- Pitch of the slider is mainly caused by loss of lift when the PM is moved away from the initial equilibrium position. It can be significantly reduced by increasing the distance between the superconductors, improving the vertical stiffness (for example, by reducing HFC) or reducing the mass of the slider.
- Pitch can also be reduced by magnetic self-alignment between the coils and the PM. This can be achieved by elevating the coils.
- A reduction of around 25% can be achieved by elevating the coils by 8 mm.

The results presented show that the contactless linear slider based on superconducting magnetic levitation technology can be very suitable for applications requiring a long stroke positioning with high precision, clean-room operation, minimised run-outs, cryogenic temperature operation and low power consumption, as demanded in certain applications for space instrumentation, medical or semiconductor manufacturing industries.

In addition, thanks to the design rules provided in this paper, similar mechanisms can be designed with the requirements of different applications.

**Acknowledgements**

The authors would like to thank to LIDAX for their technical support and cooperation in this project. We would especially like to thank Javier Serrano, Fernando Romera, Heribert Argelaguet-Vilaseca and David González-de-María for their invaluable support. This work has been partially funded by Dirección General de Economía, Estadística e Innovación Tecnológica, Consejería de Economía y Hacienda, Comunidad de Madrid, ref. 12/09.